\documentclass[preprint,aps,showpacs]{revtex4}
\usepackage{amssymb}
\usepackage{epsfig}
\usepackage{dcolumn}
\usepackage{bm}

\begin{document}

\title{Correlated metals and the LDA+U method}
\author{I. I. Mazin}
\author{A. G. Petukhov}
\altaffiliation[Permanent address: ]{Dept.~of Physics, South Dakota School of Mines and Technology, Rapid City,
SD 57701}
\affiliation{Center for Computational Materials Science, Naval Research Laboratory,
Washington, DC 20375}
\author{L. Chioncel}
\author{A. I. Lichtenstein}
\affiliation{University of Nijmegen, NL-6525 ED Nijmegen, The Netherlands}
\date{\today}

\begin{abstract}
While LDA+U method is well established for strongly correlated materials 
with well localized orbitals, its application to weakly correlated metals 
is  questionable. By extending the LDA Stoner approach onto LDA+U, we show 
that LDA+U enhances the Stoner factor, while reducing the density of 
states. Arguably the most important correlation effects in metals, 
fluctuation-induced mass renormalization and suppression of the Stoner 
factor, are missing from LDA+U. On the other hand, for {\it moderately} 
correlated metals LDA+U may be useful. With this in mind, we derive a new 
version of LDA+U that is consistent with the Hohenberg-Kohn theorem and can 
be formulated as a constrained density functional theory. We illustrate all 
of the above on concrete examples, including the controversial case of 
magnetism in FeAl.
\end{abstract}

\maketitle

One of the most influential, from practical point of view, developments in
the Density Functional Theory (DFT) in the last two decades was the LDA+U
method (see, \textit{e.g.,} Ref.~\cite{LDA+U}). This method includes the
orbital dependence of the self-energy operators, missing from the Kohn-Sham
potential, in a relatively crude, pseudo-atomic way, neglecting the fine
details of the spatial variations of the Coulomb potential. On the contrary,
the standard Local Density Approximation, LDA % (for the purpose of
%this paper the difference between LDA and gradient-corrected functionals is
%unimportant), 
accounts for the spatial variation of the Hartree potential exceedingly
well, but neglects the orbital dependence of the Coulomb interaction. 
%Importantly,
%both approaches are static; dynamic fluctuations are partially
%accounted for in the dynamic mean field theory (DMFT), which we will discuss
%later in the paper.

There is one inherent ambiguity in the LDA+U method: In LDA, all
electron-electron interactions have already been taken into account in a
mean field way. The Hubbard Hamiltonian that represents the underlying
physics of the LDA+U method also incorporates a large part of the total
Coulomb energy of the system. % which, in fact, had already been taken
%into account in a mean-field approximation
%within LDA. 
Simple combination of the LDA and Hubbard Hamiltonian thus leads to a double
counting (DC) of the Coulomb energy, so one may want to identify those parts
of the DFT expression for the total energy that correspond to the
interaction included in the Hubbard Hamiltonian and subtract them. However,
since the DFT Hamiltonian is written in terms of the total density, and the
Hubbard Hamiltonian in the orbital representation, one cannot build a direct
link between the two. Second, even if it were possible, that would be
undesirable. Spatial variation of the Hartree and the exchange-correlation
potentials is very important. It would be unreasonable to subtract that out
just because it has been already taken into account elsewhere 
in a primitive way (roughly
speaking, $UN^{2}/2)$. Rather, one wants to identify the mean-field part of
the Hubbard Hamiltonian, and subtract \textit{that}, leaving only a 
\textit{correction }to the LDA-type mean field solution.

This is not a uniquely defined procedure. Several recipes exist, and it has
been appreciated lately \cite{Novak} that the results of LDA+U calculations
may depend crucially on the choice of the DC recipe. It should be noticed
that while for strongly correlated systems the LDA+U ideology is at least
practically established, in a relatively new area of applying LDA+U to
moderately-correlated, metallic systems \cite{Shick01,Novak,Yang}, the
situation is very far from clear.

In this Letter we analyze the effect of different DC prescription on
the LDA+U results in correlated metals.
 We also present a systematic
approach to the DC problem, of which the existing recipes are special cases. 
Finally, we discuss which
problems associated with this class of materials can, in principle, be
solved within LDA+U, and which cannot.

We use for our analysis the spherically averaged form of the
rotationally-invariant LDA+U \cite{Liechtenstein}, due to Dudarev \textit{et
al.} \cite{Dudarev}: 
\begin{eqnarray}
\Delta H_{LDA+U}^{0} =\frac{U}{2}\!\!\sum_{m\sigma \neq m^{\prime }\sigma
^{\prime }}\!\!n_{m\sigma }n_{m^{\prime }\sigma ^{\prime }}\!-\!\frac{J}{2}%
\!\!\sum_{m\neq m^{\prime },\sigma }\!\!n_{m\sigma }n_{m^{\prime }\sigma } 
\nonumber \\
=\frac{1}{2}UN^{2}-\frac{1}{2}J\sum_{\sigma }N_{\sigma }^{2}-\frac{1}{2}%
(U-J)\sum_{\sigma }\mathrm{Tr}(\rho ^{\sigma }\cdot \rho ^{\sigma })
\label{U0-}
\end{eqnarray}%
where $U$ and $J$ are spherically averaged Hubbard repulsion and intraatomic
exchange for electrons with the given angular momentum $l$, $n_{m\sigma }$
is the occupation number of the $m$-th orbital, $\sigma =\pm 1$ is the spin
index, and the superscript $0$ means that the double counting terms have not
been subtracted yet. Here $\rho _{mm^{\prime }}^{\sigma }$ is the orbital
occupation matrix, $N_{\sigma }=\mathrm{Tr}(\rho ^{\sigma })$ and $%
N=\sum_{\sigma }N_{\sigma }$.

To subtract from Eq.~(\ref{U0-}) the DC term, one naturally starts with 
%It was understood from the first days of the LDA+U that it is unfeasible and
%not necessarily desirable to subtract the double counting terms from the LDA
%part of the LDA+U Hamiltonian. Instead, one has to correct Eq.~(\ref{U0-}) 
% \cite{Anisimov1,Anisimov2}. 
%subtract from the  LDA+U Hamiltonian the part of the LDA energy
%that is represented by the added Hubbard Hamiltonian\cite%
%. 
%Since the Hartree interaction is in fact better described in LDA than in the
%Hubbard Hamiltonian, the first natural step is to subtract from the Eq.~(\ref {U0-})
the first two terms in Eq.~(\ref{U0-}), $i.e.,$ the Hartree and the Stoner
energies. %\begin{equation}
%$E_{LDA+U}^{Hartree}={U}N^{2}/2$. %\end{equation}%
Both are explicit functionals of the spin density, and are likely to be
better described by LDA. % than by the Hubbard Hamiltonian, 
%so one can subtract it as well. 
To identify the DC part of the last term of the Eq.~(\ref{U0-}), which
explicitely depends on $n_{m\sigma },$ is less trivial; one needs to work
out a \textquotedblleft mean field\textquotedblright\ approximation to this
term, that is, substitute $\mathrm{Tr}(\rho ^{\sigma }\cdot \rho ^{\sigma })$
by some quantity $x_{\sigma }$ that depends solely on total spin density. Czy%
\.{z}yk and Sawatzky \cite{ChyzhikSawatsky} %\textit{et al} 
suggested that $x_{\sigma }$ should be equal to $\mathrm{Tr}(\rho ^{\sigma
}\cdot \rho ^{\sigma })$ in the limit of the uniform occupancy, $\rho
_{mm^{\prime }}^{\sigma ,LDA}=\delta _{mm^{\prime }}n_{\sigma }$, and,
consequently, $x_{\sigma }=(2l+1)n_{\sigma }^{2}$, 
%writing down the corresponding DC term as 
%\begin{equation}  \label{edc}
%E_{D.C.}=\frac{1}{2}({U}-{J})(2l+1)\sum_{\sigma }n{_{\sigma }^{2}},
%\end{equation}
where $n_{\sigma }=N_{\sigma }/(2l+1)$. This leads to the following
corrections to the total energy and the effective potential:%
\begin{eqnarray}
\Delta E_{LDA+U}^{AMF} &=&-\frac{U-J}{2}\sum_{\sigma }\mathrm{Tr}%
(\delta\rho ^{\sigma }\cdot\delta \rho ^{\sigma })%-(2l+1)n_{\sigma }^{2}\right) 
\nonumber \\
\Delta V_{LDA+U}^{AMF}(mm^{\prime }\sigma ) &=&-(U-J)\left( \rho
_{mm^{\prime }}^{\sigma }-n_{\sigma }\delta_{mm^\prime}\right) .  \label{AMF}
\end{eqnarray}%
Here AMF stands for \textquotedblleft Around Mean Field\textquotedblright\ 
\cite{ChyzhikSawatsky} 
and $\delta\rho_{mm^\prime}^\sigma =\rho_{mm^\prime}^\sigma-n_\sigma\delta_{mm^\prime}$.

For strongly correlated systems the limit of the uniform occupancy is not
correct (in fact, it is not correct even in weakly correlated systems, due
to the crystal field). %; it is easy to show that AMF therefore 
%leads to an artificial enhancement of the crystal field splitting).
Thus, it is not surprising that for the systems with strongly localized
electrons the AMF functional leads to rather unrealistic results. 
%behaves somewhat pathologically. 
%Consider, for example, Gd metal, with 7
%occupied spin-up and 7 empty spin-down orbitals. The potential (\ref{AMF})
%has zero effect, as in this case $n_{m\sigma }=n_{\sigma }.$ In reality, of
%course, the spin-up and spin-down bands form the lower and the upper Hubbard
%bands, respectively, and should be separated by a gap of the order of $U$.
This observation led \cite{ChyzhikSawatsky,Anisimov2} to another
prescription, $x_{\sigma }=(2l+1)n_{\sigma }$, 
\begin{eqnarray}
\Delta E_{LDA+U}^{FLL} =-\frac{U-J}{2}\sum_{\sigma }\left( \mathrm{Tr}%
(\rho ^{\sigma }\cdot \rho ^{\sigma })-(2l+1)n_{\sigma }\right)  \nonumber \\
\Delta V_{LDA+U}^{FLL}(mm^{\prime }\sigma ) =-(U-J)\left( \rho
_{mm^{\prime }}^{\sigma }-\frac{1}{2}\delta_{mm^\prime}\right) ,  \label{FLL}
\end{eqnarray}%
which produces the correct behavior in the fully localized limit (FLL) where 
$n_{m\sigma }=0$ or 1. Most of the modern LDA+U calculations utilize one of
these two functionals, although in real materials the occupation numbers lie
between these two limits.

In the AMF the LDA+U correction to the electronic potential, Eq.~%
\ref{AMF}, averaged over all occupied states, is zero. This is a possible
way to define a mean field (cf. the Slater approximation to the Fock
potential), but not the way used in the DFT. The latter is a mean field
theory that produces the correct total energy, not the correct average
potential. AMF and FLL represent the \textquotedblleft
DFT\textquotedblright\ mean field if all occupation numbers are all the
same, or are all 0 or 1, respectively. It is easy to show that $%
(2l+1)n_{\sigma }^{2}\leq \mathrm{Tr}(\rho ^{\sigma }\cdot \rho ^{\sigma
})\leq (2l+1)n_{\sigma }$, so that AMF always gives a negative, and FLL a
positive correction to the total energy, while the right (in the DFT sense)
recipe should give zero correction to the total energy.  
That can be achieved by using a
linear interpolation between the two extremes corresponding to AMF and FLL, $%
x_{\sigma }=(2l+1)\left( \alpha n_{\sigma }+(1-\alpha )n_{\sigma
}^{2}\right)$, where $0\leq \alpha\leq 1$, and
\begin{displaymath}
\Delta E_{LDA+U}^{DFT}\!\!=\!\!-\frac{{U}\!\!-\!\!{J}}{2}\sum_{\sigma }\left[\mathrm{Tr}
(\delta\rho^\sigma\!\!\cdot\delta\rho^\sigma)
\!\!-\!\!(2l\!+\!1)\alpha n_\sigma(1-n_\sigma)\right] 
\end{displaymath}%\nonumber\\
\begin{equation}
\Delta V_{LDA+U}^{DFT}(mm^{\prime }\sigma )\!=\!-({U}\!-\!{J})\!\!
\left[\rho_{mm^{\prime }}^{\sigma }\!-\!\left((1\!\!-\!\!\alpha)n_{\sigma }\!+\!\frac{\alpha}{2}
\right)\delta_{mm^\prime}\right]\!.  
\label{MP}
\end{equation}
In the spirit of the DFT, $\Delta E_{LDA+U}^{DFT}\!=\!0,$ so
\begin{equation}
%0\leq 
\alpha=\frac{\sum_\sigma\mathrm{Tr}(\delta \rho ^{\sigma }\cdot \delta
\rho ^{\sigma })}{(2l+1)\sum_\sigma n_{\sigma }(1-n_{\sigma })}
%\leq 1.  
\label{alpha}
\end{equation}

We emphasize that $\alpha $ is not adjustable, nor is it a formal functional
of the charge density, but it is a material-dependent
parameter (like, say, $U$
itself), defined by the \textit{self-consistent} occupation matrix. However,
in practical calculations it is better to recompute $\alpha $ after each
iteration, as the current value of $\rho _{mm^{\prime }}^{\sigma } $
changes. Note that the total energy is given by the regular LDA expression
that only implicitly depends on $U$ and $J$ \textit{via} the changing
density distribution; it is variational with respect to the charge density
at a fixed $\alpha $, but \textit{not} variational with respect to $\alpha 
$ itself. The fact that this prescription is derived according
to the DFT ideology allows one to formulate the proposed
LDA+U functional (unlike the existing LDA+U
functionals) as a \emph{constrained DFT theory} \cite{Dederichs} at a given $%
\alpha$, with the constraint given by Eq.~(\ref{alpha}).
$(U-J)/2$
appears then as a Lagrange multiplier. %This point of view
%has several important consequences. First of all, it is obvious that
%constraint (\ref{x}) is not unique simply because
%one can envision many other ways of subtracting DC term
%that also can be presented in constrained LDA form. 
%However, our approach allows us
%to select between different functionals systematically by giving the
%preference to less restrictive constraints. Second, one can always
%perform calculations with fixed $\alpha$ rather than with fixed $U_{eff}=
%U - J$ and then impose Eq. (\ref{x}) to find
%$U_{eff}$. In some cases this procedure is more
%informative than that of with fixed $U_{eff}$. 
%We emphasize that neither of the limiting cases
%such as AMF ($\alpha=0$) or FLL ($\alpha=1$) can be treated as a constrained
%LDA approach  because Eq. (\ref{x}) with $\alpha$ = 0 or 1 cannot be
%satisfied for any realistic solid state system with hybridized local orbitals. 

\begin{figure}[tbph]
\epsfig{file=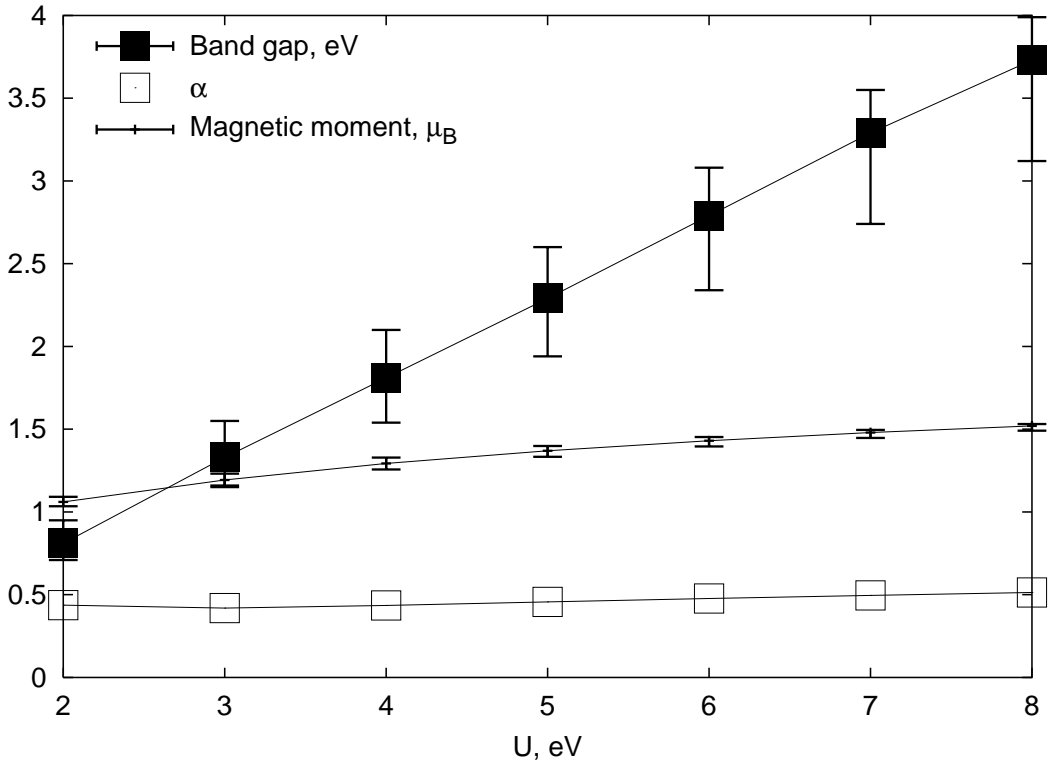,width=.9\linewidth,clip=true}
\caption{Mott-Hubbard band gaps, and magnetic moments of antiferromagnetic
NiO for three flafors of LDA+U. The upper and lower values of the
\textquotedblleft error bars\textquotedblright\ correspond to the FLL and
AMF functionals, respectively.}
\label{NiO}
\end{figure}

We have tested the proposed functional (\ref{MP}) %in some other
%systems such as Mott-Hubbard insualtors and 4f metals.
%As a first example we will consider electronic structure of 
on NiO, a prototypical compound for LDA+U calculations (see, e.g., \cite%
{Alluani}). Fig. \ref{NiO} shows the band gap and the mangetic moment of NiO
as a function of $U$ at $J=0.95$ eV for three different functionals (Eqs. %
\ref{AMF}, \ref{FLL}, and \ref{MP}), calculated within the linear-muffin-tin
orbital (LMTO) method in the atomic sphere approximation (ASA). 
%The parameter $\alpha$ of the functional (\ref{MP}) is shown as well.
%The upper and lower values of the ``error bars'' correspond to the FLL
%and AMF functionals respectively. 
The parameter $\alpha \simeq 0.5$ is almost independent on $U$. Accordingly,
the results of our calculations based on Eq.~ \ref{MP} for both band gaps
and magnetic moments lie right between those for AMF and FLL
calculations, and the effect of $U$ is reduced compared to the FLL
calculations. This is in accord with a known observation \cite{Alluani} that
in NiO the FLL LDA+U gives the best agreement with the experiment for $%
U\lesssim $6 eV, smaller than $U\simeq $ 8 eV calculated from the first
principles\cite{Alluani,Gunnarson}. 

Our next example is a weakly correlated metal FeAl. This paramagnetic
material has attracted attention due to a recent suggestion by Mohn et al. 
\cite{Novak} that the short-range Coulomb correlations within the LDA+U may
be responsible for suppression of ferromagnetism found in all LDA calculations.
More specifically, they found in their AMF LDA+U calculations a reduction of
the density of states (DOS) at the Fermi level, $D_{F},$ which was
sufficient to make the Stoner criterion smaller than 1 and stabilize the
paramagnetic state. To analyze this result, it is important to revisit the
Stoner theory for the LDA+U case.

In DFT, the Stoner parameter $I$ is defined as $I=-2\partial
^{2}E_{xc}/\partial M^{2}$, the second derivative of the exchage-correlation
energy with respect to the total magnetic moment. The paramagnetic ground
state is unstable when $D_{F}I\geq 1.$ This can be derived from the force
theorem, which states that the total energy for small magnetizations can be
computed by assuming a rigid shift of the bands by $b=\pm M/2D_{F},$ so
that the gain in the interaction energy, $-IM^{2}/4$, competes with the loss
in the one-electron energy, $M^{2}/4D_{F}.$ In the LDA+U the criterion
holds, but the product $D_{F}I$ changes, not only because $D_{F}$ changes,
but also because the newly added interaction energy depends on $M$. Indeed,
the force theorem calls for a change $\delta \rho _{mm^{\prime }}^{\sigma
}=b\sigma D_{mm^{\prime }}$, where $D_{mm^{\prime }}=-\pi ^{-1}\mathrm{Im}%
~G_{mm^{\prime }}(E_{F}).$ When applied to the functionals Eq.~(\ref{AMF}) 
- Eq.~(\ref{MP}), it generates a change in the interaction energy, which
results in an additional contribution to the Stoner parameter, 
\begin{equation}
\Delta I(\alpha )=(U-J)\left( {\mathrm{Tr}}(D\cdot D)-\frac{(1-\alpha )({%
\mathrm{Tr}D})^{2}}{(2l+1)D_{F}^{2}}\right)  \label{I}
\end{equation}

In the limit of the uniform occupancy, Eq.~(\ref{I}) for the FLL case ($\alpha
=1$) reduces to $(U-J)/(2l+1)$. Given that the LDA Stoner parameter, $I,$ is
of the same order as $J,$ we obtain for the total Stoner parameter $%
I_{FLL}\approx (U+2lJ)/(2l+1),$ which the well known expression for the
Stoner factor in the atomic Hubbard model. On the contrary, $\Delta I_{AMF}$
($\alpha =0$) in this limit is zero. In real metals $D_{mm^{\prime }}$ is
complicated due to crystal field effects. Let us consider, for illustration,
d-electrons in a cubic environment, and introduce the difference $\Delta
D=D_{eg}-D_{t2g}$, where $D_{eg}$ and $D_{t2g}$ are $e_{g}$ and $t_{2g}$ DOS
per orbital at $E_{F}$, as a measure of the crystal field. This gives rise
to a %(quadratic in $\Delta D$) 
contribution to $\Delta I_{AMF}=\frac{5}{24}(U-J)(\Delta D/D_{F})^{2}$.
However, when LDA+U reduces $D_{F}$, and $\Delta I_{AMF}$ is not large
enough to overcome the decrease in $D_{F}$, LDA+U may stabilize the
paramagnetic state(cf.  Ref. \cite{dimitri}), as, for instance, observed in
a very narrow range of large $U$'s for FeAl by Mohn \textit{et al.} \cite%
{Novak} (of course, only in the AMF functional; the FLL functional produces
a large $\Delta I\approx (U-J)/5$, always increasing the tendency to
magnetism).

\begin{figure}[tbph]
\epsfig{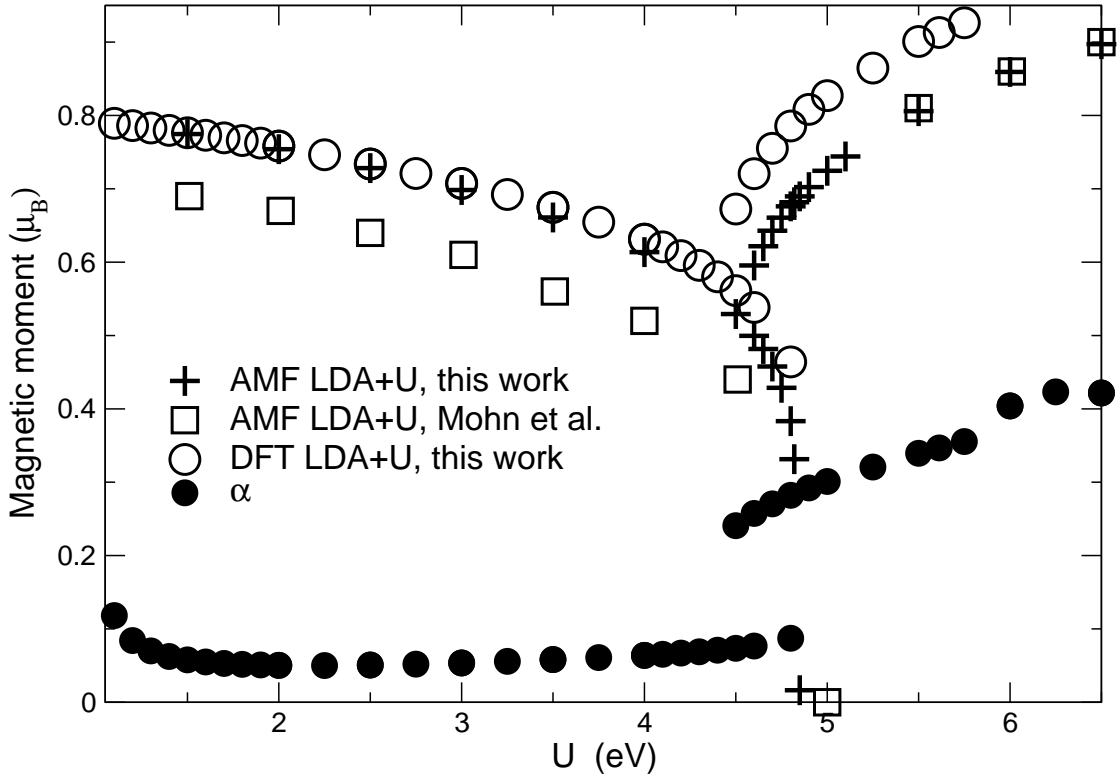}
\caption{Magnetic moments of FeAl for AMF and DFT flavors of LDA+U compared
with the results of Mohn et al. (Ref.~{\protect\cite{Novak}})}
\label{FeAl}
\end{figure}

With this in mind, we performed LMTO-ASA calculations for all three LDA+U
functionals, using fixed $J=0.95$ eV. The results for $U$-dependence of the
magnetic moment and $\alpha $ are shown in Fig.~\ref{FeAl} and compared with
those by Mohn et al. \cite{Novak}. In our AMF calculations we also found a
paramagnetic solution for $U=$ 4.85 eV, which however coexsits with a
ferromagnetic high spin solution (Fig.~\ref{FeAl}). Note that for well
localized orbitals there is no difference whether the $(U-J)$ term is
applied inside the atomic sphere or only inside the MT sphere, as in Ref.~%
\cite{Novak}; however, in less localized cases, where a noticeable part of
the $d$-orbitals spills out of the MT sphere, the effect of the same $U$ is
smaller when applied only inside the MT sphere. One can see in Fig.~\ref%
{FeAl} that, indeed, our calculations with large $U$ yield large $\alpha $'s
and agree very well with Ref. \cite{Novak}, while for small $U$ (small $%
\alpha $) the effect of $U$ in our ASA calculations is stronger than in Ref. 
\cite{Novak}. 

All LDA+U functionals shift unoccupied bands up and occupied bands down.
Therefore LDA+U broadens the bands crossing the Fermi level. Because of this
broadening, in FeAl for small $U$  the parameter $\alpha $ is initially
decreasing (Fig.~\ref{FeAl}) with a minimum $\alpha =0.05$ at $U=2$ eV. The
magnetic moment also decreases in this region. At larger $U,$ $\alpha $
starts growing again. At this point it is instructive to apply the logic of
the constrained LDA approach in which for every fixed $\alpha $ the total
LDA energy is minimized under the constraint $\sum_{\sigma }\mathrm{Tr}%
(\delta \rho ^{\sigma }\cdot \delta \rho ^{\sigma })/[(2l+1)\sum_{\sigma
}n_{\sigma }(1-n_{\sigma })]=\alpha $, $(U-J)/2$ being the Lagrange
multiplier. For $\alpha \leq 0.087$ (Fig.~\ref{FeAl}) of the two possible
solutions with $U<2$ eV and $U>2$ eV we should choose the one with lower
energy (smaller $U).$ As a result, we find two admissible domains for $U:$
an AMF-like with $U<2$ eV and a FLL-like with $U\gtrsim 5$ eV. The latter is
clearly unphysical. Both solutions are ferromagnetic. The solutions with
intermediate values of $U$ and reduced magnetic moments are inadmissible in
the framework of the constraint DFT formulation. 

On the contrary, our explanation of the paramagnetism in FeAl is that the
ferromagnetism instability is suppressed by the critical spin fluctuations.
There are many other systems for which the fluctuations in the vicinity of a
quantum critical point reduce the tendency to magnetism. Further examples
include Sr$_{3}$Ru$_{2}$O$_{7}$ ($M_{LDA}\approx 0.8$ $\mu _{B},$ $M_{\exp
}=0),$ ZrZn$_{2}$ ($M_{LDA}\approx 0.7$ $\mu _{B},$ $M_{\exp }=0.2$ $\mu
_{B}),$ and other.  The physics that is missing from both LDA and LDA+U
equations in such systems can be described as exchange of virtual electronic
excitations, roughly speaking, plasmons or (para)magnons. This leads to
\textquotedblleft dressing\textquotedblright\ of the one-particle
excitations in the same way as the electron-phonon coupling
\textquotedblleft dresses\textquotedblright\ electrons near the Fermi
surface, although in a correlated metal such mass renormalization effects
occur on a large energy scale (of the order of $U$ or $J).$ LDA calculations
cannot reproduce such a dressing, which has been observed in many different
ways experimentally. For instance, LDA calculations do not explain large
mass renormalizations in Sr$_{2}$RuO$_{4}$ \cite{mackenzie}, and large
specific heat renormalization in many correlated metals, produce too large
plasma frequencies, $e.g.,$ in YBa$_{2}$Cu$_{3}$O$_{7},$ yield an optical
absorption spectrum in CrO$_{2}$ shifted by about 20\% to higher frequency,
as compared with experiment \cite{Claudia}, and overestimate the exchange
splitting in Ni by a factor of 2 \cite{NiPES}. In all these cases 
%correlation effects lead to \textit{narrowing} of the
%valence bands, 
the total width of the $d$-bands is \textit{decreased}, as opposed to 
\textit{broadening} inherent to LDA+U. Here the essential physics is missing
from the LDA+U as well as in LDA, while the spatial variation of the
mean-field Coulomb interaction is treated better by the LDA. The missing
physics is associated, to a large degree, with dynamic fluctuations. 

\begin{figure}[t]
\epsfig{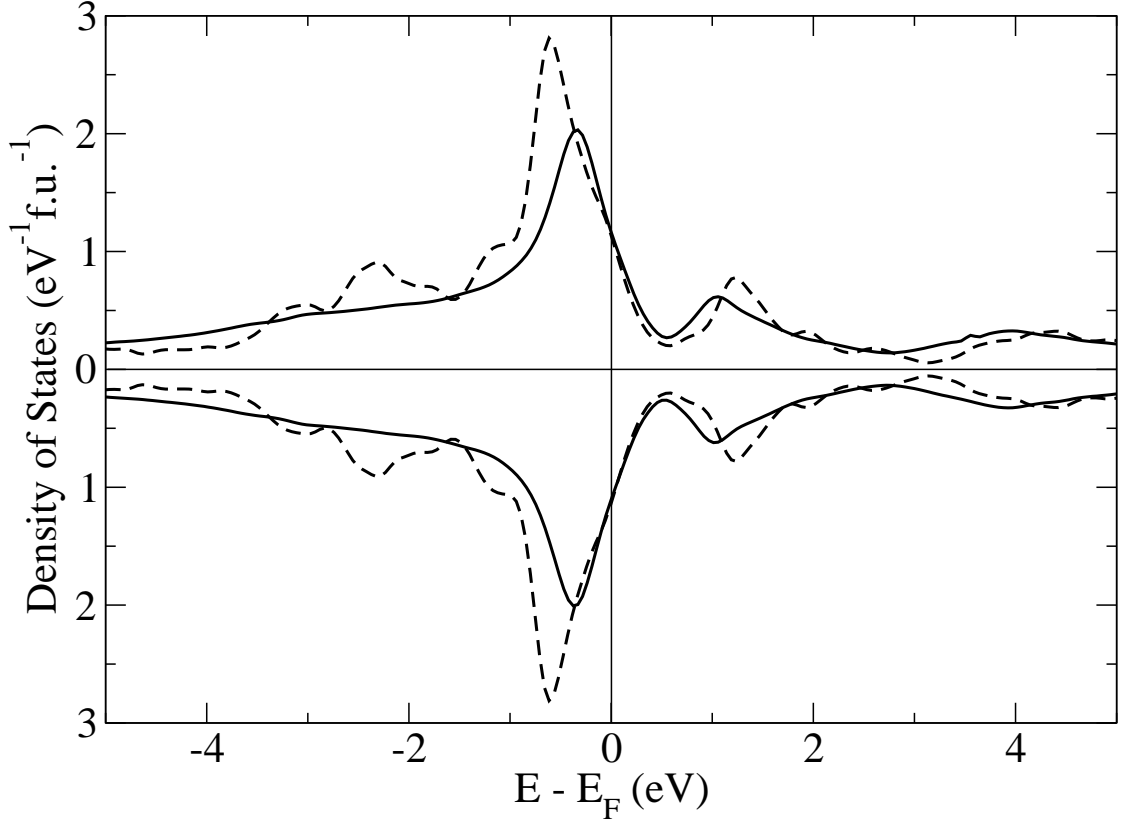}
\caption{FeAl density of states, $D(E)$, in DMFT (solid line) compared with
the nonmagnetic LDA. The DMFT solution is stable, the LDA is not (a
ferromagnetic solution is stable), despite the same $D(E_{F})$.}
\label{lulu}
\end{figure}

The dynamic version of the LDA+U method, %the Dynamic Mean Field Theory
the Dynamic Mean Field Theory (DMFT) \cite{Georges}, which can account for
some spin fluctuations \cite{spflex}, resolves many of these problems. For
instance, the  mass renormalization in Sr$_{2}$RuO$_{4}$ is 3-4 \cite%
{mackenzie}, far greater than possible renormalization due to the phonons.
We applied all three flavor of LDA+U to Sr$_{2}$RuO$_{4}$ and found no mass
renormalization compared to LDA. On the other hand, Eliashberg-type
calculations \cite{mazinsingh} of the renormalization due to spin
fluctuations, using a spectrum deduced from the LDA band structure, give
mass renormalizations of the right order. Similarly, DMFT   explicitely
narrows the bands in Sr$_{2}$RuO$_{4}$ and enhances the electronic mass \cite%
{LL}.
With this in mind,  
we applied the DMFT with a realistic $U=2$ eV to FeAl and found
the paramagnetic state to be perfectly stable, whith the bands \textit{%
narrower }than in LDA, and the density of states practically the same (Fig.~%
\ref{lulu}). In other words, the spin fluctuations 
%near the quantum critical point 
effectively reduce the Stoner factor $I$. %This
%reduction overcomes some increase in the density of states.

To conclude, we observe that no %neither of the two available 
LDA+U functional correctly describes the essential physics of the weakly
correlated metals: (i) reducing the band dispersion by dressing of the
one-particle excitation, and (ii) spin fluctuations near the quantum
critical point. One functional, labeled FLL here, correctly describes the
important physics in the limit of well localized electrons, and can be
recommended in this case. The other functional, labeled AMF, is exact in a
hypothetical material with the uniform orbital occupancies. Although neither
functional accounts for the fluctuation effects, LDA+U may be useful, if
applied with a grain of salt, in moderately correlated metals. For this
case, we propose a recipe that accounts for an incomplete localization and
reduces to AMF or FLL in the appropriate limits. Finally, it is worth noting
that in many correlated materials the spin-orbit interaction plays a key
role. Since our $\alpha $ does not depend on spin, this prescription can be
also formulated in terms of the full $(4l+2)\times(4l+2)$ occupation matrix $%
\rho $ and $n=\mathrm{Tr}(\rho)/(4l+2)$. 
Eq. (\ref{alpha}) should be replaced with $\mathrm{Tr}(\delta \rho \cdot
\delta \rho )=$ $(4l+2)\alpha n(1-n)$. This formulation has another
advantage in the case of a half-filled band, like in Gd, because in this
limit it reduces to more physically meaningful in this case FLL, rather than
to AMF as the nonrelativistic Eq.~(\ref{MP}).

%We acknowledge many useful discussions with S. C. Hellberg and D. A.
%Papaconstantopoulos. 
This work is supported in part by ONR and by NSF (Grant No DMR-0071823).

%$n(E_F)=\frac{1}{3\pi^2}\left(\frac{2m}{\hbar^2}E_F\right)^{3/2}$

%$n(E_F)=\frac{1}{6\pi^2}\left(\frac{2m}{\hbar^2}\right)^{3/2}
%\left[(E_F)^{3/2}+(E_F-\Delta)^{3/2}\right]$

%\bibliography{/home/main5/petukhov/tex/STONER/DRAFT/stoner}
\bibliographystyle{apsrev}
\bibliography{stoner}

\end{document}